\documentclass{aastex}          
\usepackage{spr-astr-addons}    

\begin{document}
%
\title{The Fluxgate-Searchcoil Merged (FSM) Magnetic Field Data Product for MMS}

\shorttitle{The Fluxgate-Searchcoil Merged Data Product}
\shortauthors{Argall et al.}

\author{M. R. Argall\altaffilmark{1}}
\affil{University of New Hampshire, Durham, NH, USA}
\email{matthew.argall@unh.edu}
\and
\author{D. Fischer\altaffilmark{2}}
\affil{Space Research Institute, Austrian Academy of Sciences, Graz, Austria}
\and
\author{O. Le Contel\altaffilmark{3}}
\and
\author{L. Mirioni\altaffilmark{3}}x
\affil{Laboratory for Plasma Physics, Paris, France}
\and
\author{R. B.  Torbert\altaffilmark{1}}
\and
\author{I. Dors\altaffilmark{1}}
\and
\author{M. Chutter\altaffilmark{1}}
\and
\author{J. Needell\altaffilmark{1}}
\affil{University of New Hampshire, Durham, NH, USA}
\and
\author{R. Strangeway\altaffilmark{4}}
\affil{University of California, Los Angeles, CA, USA}
\and
\author{W. Magnes\altaffilmark{2}}
\affil{Space Research Institute, Austrian Academy of Sciences, Graz, Austria}
\and
\author{C. T. Russell\altaffilmark{4}}
\affil{University of California, Los Angeles, CA, USA}

\altaffiltext{1}{University of New Hampshire, Durham, NH, USA}
\altaffiltext{2}{Space Research Institute, Austrian Academy of Sciences, Graz, Austria}
\altaffiltext{3}{Laboratory for Plasma Physics, Paris, France}
\altaffiltext{4}{University of California, Los Angeles, CA, USA}

\begin{abstract}
The Fluxgate-Searchcoil Merged (FSM) data product for the Magnetospheric Multiscale (MMS) mission is created by combining the level-2 science quality data from the fluxgate and searchcoil magnetometers  into a single level-3 data product. The merging method involves noise floor and calibration parameters determined both on the pre- and post-flight. Here, we describe the statistical inter-calibration process as well as the merging filter.
\end{abstract}


%
\section{Introduction}
\label{sec:intro}
The Magnetospheric Multiscale (MMS) mission \citep{Burch:2015} consists of four identically instrumented satellites. Within the FIELDS consortium \citep{Torbert:2014} on each satellite are two fluxgate magnetometers (FGMs) \citep{Russell:2014} and a searchcoil magnetometer (SCM) \citep{LeContel:2014}. In this manuscript, we describe a method of merging the vector magnetic field data products from FGM and SCM into a single data product, named the fluxgate searchcoil merged (FSM) magnetic field.

The two FGM types are the analog fluxgate (AFG) and digital fluxgate (DFG), for which the sensors differ by the addition of a capacitor in AFG. The electronics, however, were designed based on different operating principles \citep{Russell:2014, Magnes:2008}, providing robust redundancy.


DFG data is decimated from its raw sapling rate of 8192\,S/s by two different methods, named DEC32 and DEC64. In DEC32 mode, the sampling rate is reduced to 256\,S/s via a 32-point average, then is decimated to 128\,Hz. This mode produces aliasing, but reduces the group delay so that the magnetic field can be piped to the Electron Drift Instrument (EDI) \citep{Torbert:2015} faster, reducing the amount of prediction required. In DEC64 mode, the sampling rate is reduced to 128\,S/s directly via 64-point averaging. Each mode has unique effects on the measured signal and so are handled separately. In the analysis that follows, we focus on DEC32, but the same procedure has been applied to DEC64.




A previous iteration of the merging method presented in this article was applied to Cluster data to study current density associated with magnetic reconnection using the multi- and single-spacecraft techniques \citep{Argall:2014}. For that implementation, data was merged in the frequency domain with a step function. The implementation outlined below makes use of a finite impulse response (FIR) filter constructed from windowed, truncated $sinc$ functions.

Merged fluxgate and searchcoil datasets have been created in the past for scientific investigations.
 
 \textbf{* Motivation}

\section{Method}
\label{sec:method}

\subsection{Data Preparation}
\label{sec:DataPrep}
Before merging can take place, FGM and SCM undergo the same calibration process as the publicly available level 2 science quality data. Ground calibrations for each instrument were performed. A test signal driven by a spectrum analyzer
is swept through the operational frequency range and compared to the output signal measured by the magnetometer. The resulting transfer function is then convolved into measurements taken in-flight to correct gain as a function of frequency. However, because the transfer functions for AFG and DFG are flat up to \textbf{XX\,Hz}, they are not applied prior to undergoing the in-flight calibration process.

\begin{figure}[tb]
\includegraphics[width=\columnwidth]{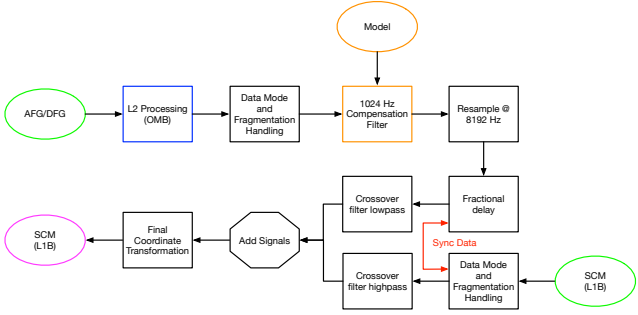}
\caption{FSM data processing workflow.} 
 \label{fig:method}
 \end{figure}

For FGM, the in-flight calibration process entails 1) spin tone removal to determine orthogonalization parameters, 2) range joining to remove jumps between low- and high-range, 3) cross-calibration with the Electron Drift Instrument (EDI) \citep{Torbert:2014} to remove spin-axis offsets, 4) earth-field comparisons to determine absolute calibration factors, and 5) inter-spacecraft comparisons to remove relative differences \citep{Russell:2014}. In addition, corrections due to temperature drifts are taken into account (\textbf{Bromund, et al., 2018}).


The in-flight calibration process for SCM consists of 1) applying the transfer function, with corrections obtained via an on-board test signal; 2) high pass filtering with a cut-off frequency at the lower operational range (1\,Hz in burst mode); and 3) high frequency corrections \citep{LeContel:2014}.

Finally, \citep{Fischer:2016}.



\subsection{Merging Interval}
\label{sec:MergeInterval}
After the data has been process to level 2 and the compensation filters have been applied, the next step is to determine an appropriate frequency or frequency range over which the FGM and SCM datasets should be merged. This is done by obtaining a statistical representation of the in-situ noise floor. Figure~\ref{fig:method} outlines the procedure. The top row of panels show the power spectral density (PSD) of the x-component of the magnetic field measured by DFG and SCM for an entire burst data file on 2015-09-01. Here and in the figures that follow, DFG data is presented in the left column and SCM data in the right column. The second row shows the distribution of signal as a function of frequency. At each frequency, the PSD was separated into bins of size $0.5\,Log(\text{nT}^{2}\text{/Hz})$ and accumulated over the duration of the burst file. In the last row, histograms from all burst files in September of 2015 were accumulated into a single distribution. The idea behind the method is that the instruments will measure signals associated with noise more often than it detects signals driven by physical phenomena, and the noise floor can be extracted by fitting the distribution with a gaussian function.

\begin{figure}[tb]
\includegraphics[width=\columnwidth]{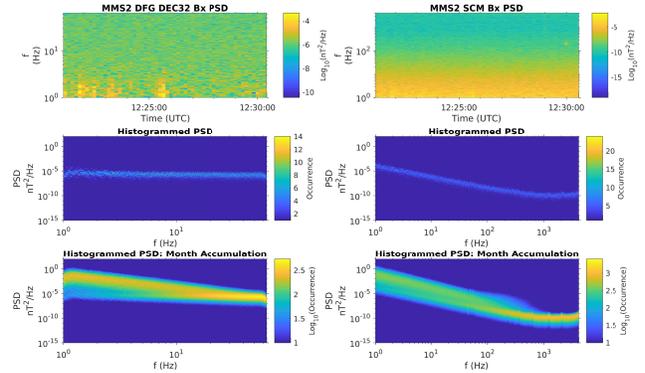}
\caption{Accumulation of signal statistics. (a) A power spectrogram from a single burst file; (b) a histogram of signal as a function of frequency and time; (c) histogram similar to (b) but accumulated over all burst files in the month.} 
 \label{fig:method}
 \end{figure}
 
To compute the PSDs, DFG is first high-pass filtered with a cut-off frequency of $f_{c} = 1$\,Hz to mimic the SCM calibration process and to prevent leaking of DC signal to higher frequencies. Next, a $sinc$-function and Chebyshev window taper $T = 20$\,s data intervals which are then Fast Fourier Transformed (FFT-ed) to the frequency domain to calculate PSD. Subsequent windows contain 50\% overlap. A time interval of $T=20$\,s was chosen for both DFG and SCM for two reasons: 1) a frequency resolution of $\Delta f = 1 / T = f_{s} / N = 0.05$\,Hz (where $f_{s}$ is the sample rate and $N$ is the number of samples per FFT) is high enough to decipher peaks in the statistical distributions, and 2) the frequency bins and their spacing are the same for DFG and SCM up to the Nyquist frequency $f_{N}$ of DFG. This allows us to make gain and phase comparisons between the two instruments, as will be shown later.

\begin{figure}[tb]
\includegraphics[width=\columnwidth]{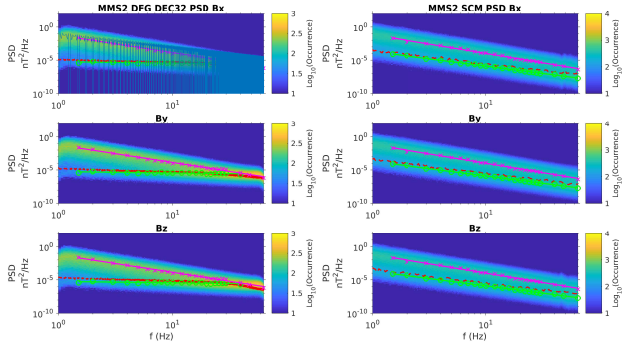}
\caption{Distribution of signal as a function of frequency over the course of one month for DFG in DEC32 (left) and SCM (right). Each row represents a component in OMB coordinates.} 
\label{fig:nfhist}
\end{figure}

\begin{figure}[tb]
\includegraphics[width=\columnwidth]{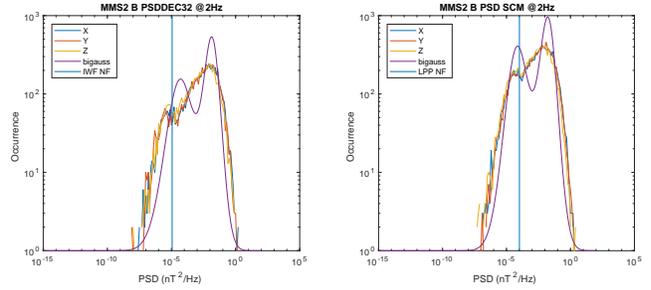}
\caption{Distribution of signal at 2\,Hz for each component of DFG (left) and SCM (right). The vertical line in each panel represents the ground-measured noise floor. In these bimodal distributions, one peak is associated with noise and the other with real signal.} 
\label{fig:2Hz}
\end{figure}

Such a procedure has been applied to all three components of the magnetic field (Figure~\ref{fig:nfhist}). The noise floor determined on the ground for DFG and SCM is represented by the magenta curve in each panel. Note that the ground-measured noise floor overlaps with the distribution present in the second row of Figure~\ref{fig:method}, but there is a second peak in the distribution at low frequencies. This is better seen in Figure~\ref{fig:2Hz}, which shows vertical cuts through each component at 2\,Hz, with the mean value of the noise floor of each component plotted as a blue vertical line. A peak at a PSD of ${\sim}10^{-6}\,\text{nT}^{2}\text{/Hz}$ for DFG and  ${\sim}10^{-5}\,\text{nT}^{2}\text{/Hz}$ occurs slightly below the value of the ground-measured noise floor at 2\,Hz. This peak is the in-situ measured noise floor.

A second, larger distribution of occurs to the right of the first. Unlike the peaks associated with the noise floor, it is observed at the same PSD value of ${\sim}10^{-2}\,\text{nT}^{2}\text{/Hz}$ for both DFG and SCM. Plots similar to Figure~\ref{fig:method} made for individual burst intervals containing significant wave activity reveal peaks in the distribution coincident with this second peak (not shown). Thus, this second peak is associated with signal driven by physical phenomena.
 
Mean noise and signal levels are extracted by fitting a bigaussian distribution to the data at each frequency (e.g. the purple trace in Figure~\ref{fig:2Hz}) \textbf{the fits are bad!}. The cross-over frequency, $f_{x}$ is then defined as the frequency at which the SCM noise floor crosses that of DFG. Data is merged with a finite impulse response (FIR) filter (see \S\ref{sec:Filter}) at the cross-over frequency, $f_{x} = 7$\,Hz
 
\section{Gain and Phase Delay}
\label{sec:GainPhase}
To calculate gain and phase delays, the niose and signal distributions must be separated. This is accomplished by selecting a frequency-dependent threshold value for PSD below which all values are considered noise and above which all values are considered signal. The threshold value is taken to be the maximum between components of the ground-measured noise floor at each frequency. This was chosen because the pre-flight noise floor is higher than the in-situ noise floor
By separating the signal from the noise distributions, we are able to compare DFG and SCM signals to determine long-term trends in gain and phase delays.

\begin{figure}[tb]
\includegraphics[width=\columnwidth]{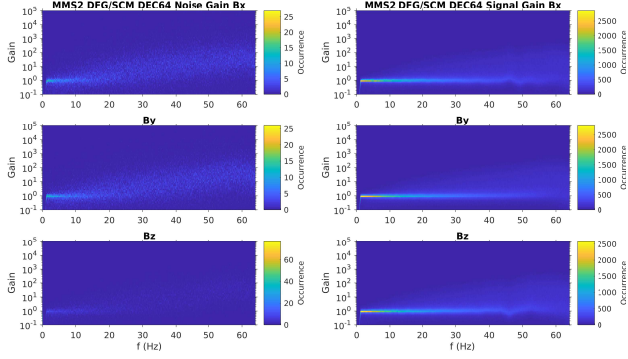}
\caption{Distribution of gain associated with noise (left) and signal (right) for each component. The gain and phase delay are measured with respect of DFG. Noise gain follows the expected gain correction between pre-flight and in-situ noise floors. Signal gain remains unity throughout the interval, despite a broadening of the distribution toward $f_{N}$.} 
\label{fig:Gain}
\end{figure}

Gain for both the noise floor (left) and signal (right) distributions is shown in Figure~\ref{fig:Gain}. It is computed as $G = \left| R^{2} \right|$, where $R = B(f) / \delta B(f)$, and $B(f)$ and $\delta B(f)$ are the FFTs of the magnetic field measured by DFG and SCM, respectively. The histogram process is the same as that described in \S\ref{sec:DataPrep} for PSD, with bin sizes for gain and frequency of 0.5 and 0.05\,Hz, respectively. Noise gain $\bar{G}_{n} < 1.0$ at low frequencies where the DFG noise floor is lower than that of FGM. Then, near 7\,Hz $\bar{G}_{n}$ increases above 1.0 as the SCM noise floor drops, further motivating our choice of cross-over frequencies. Signal gain $\bar{G}_{s} = 1.0$ throughout the frequency range of DFG, which is a testament to the calibration and cross-calibration efforts undertaken by the magnetometer teams. The width of the distribution increases toward higher frequency, which is primarily caused by the lack of sufficiently strong signal and our inability to cleanly separate noise from signal at $f > 20$\,Hz.
 
\begin{figure}[tb]
\includegraphics[width=\columnwidth]{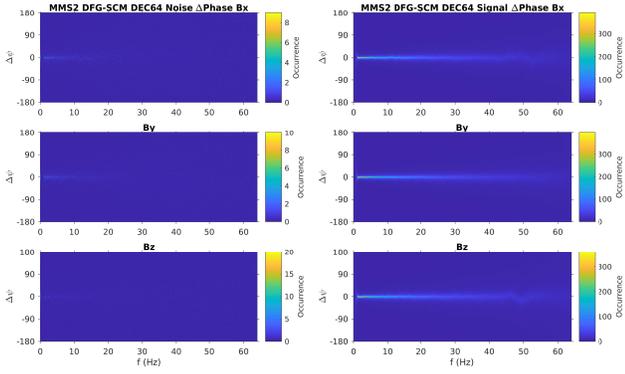}
\caption{Distribution of phase delay associated with noise (left) and signal (right) for each component. The gain and phase delay are measured with respect of DFG. Noise phase delay appears random. Signal phase delay remains zero throughout the interval, but, similar to gain, the distribution broadens toward $f_{N}$.} 
\label{fig:Phase}
\end{figure}

Phase delay distributions are shown in Figure~\ref{fig:Phase}. Phase delay is computed as $\Delta\psi = \tan^{-1} \left( \Re[R]/ \Im[R] \right)$, then histogrammed in the same manner as PSD and gain (\S\ref{sec:DataPrep}). Distributions associated with noise (left) have apparently random phase delays. Meanwhile, distributions associated with signal (right) are strongly peaked at $\Delta\psi = 0^{\circ}$, indicating the high timing accuracy between the two instruments. Again, the distribution spreads as frequency approaches $f_{N}$
 
\section{Merging Filter}
\label{sec:Filter}	
By comparing the in-situ determined noise floors, we were able to determine the cross-over frequency at which the data products should be merged. Figure~\ref{fig:Filter} shows the design of the filter. The first panel contains the impulse response of the FIR filter, constructed from three separate low-pass FIR filters with cut-off frequencies. Two with cut-off frequencies of $f_c = 4$ and 7\,Hz serve to provide an interval of 50\% gain surrounding $f_{x}$ for each instrument. They are $sinc$ functions truncated at 16385 points (2\,s duration) and tapered with a Blackman window. A third low-pass filter with $f_{c} = 32$\,Hz acts as an anti-aliasing filter. It is a $sinc$ function truncated at 2049 points (0.25\,s duration) tapered with a Chebychev window. The resulting 4.25\,s duration filter is convolved with the upsampled DFG data. A complementary high-pass is formed via spectral inversion and convolved into the SCM data.

\begin{figure}[tb]
\includegraphics[width=\columnwidth]{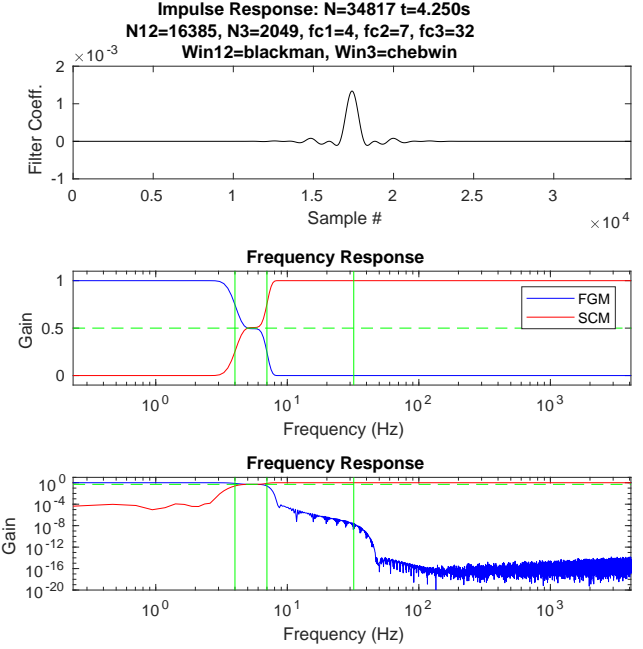}
\caption{Finite impulse response filter used to merge the DFG and SCM data products. (a) The filter coefficients of the time domain impulse response; (b) the frequency response plotted on a linear scale to demonstrate the pass-band and roll off; and (c) the frequency response plotted on a log-scale to show properties of the stop-band. Vertical green lines mark the cut-off frequencies of each filter. Blue traces are of the low-pass filter applied to DFG while red traces depict the high-pass filter applied to SCM. This sum of the blue and red traces is unity.} 
\label{fig:Filter}
\end{figure} 

Panels c and d depict the frequency response of the merging filter. Green vertical lines represent the cut-off frequencies used. The Blackman and Chebychev tapering windows result in a smooth passband with no overshoot (panel a). The Blackman window results in a 40\,dB attenuation within 1\,Hz of $f_{c}$ (panel b) and its passband is narrower than that of the Blackman window. For anti-aliasing purposes, a narrow pass-band was not as important as a strong stop-band, which is provided by the Chebychev window. The $f_{c} = 32$\,Hz low-pass filter produces an additional 100\,dB attenuation at roughly $f_{N}/2$ for DFG.

Applying these merging filters to the noise floors of DFG and SCM identified in Figure~\ref{fig:nfhist} shows the improvement in noise within the merged frequency interval of the FSM data product.

\begin{figure}[tb]
\includegraphics[width=\columnwidth]{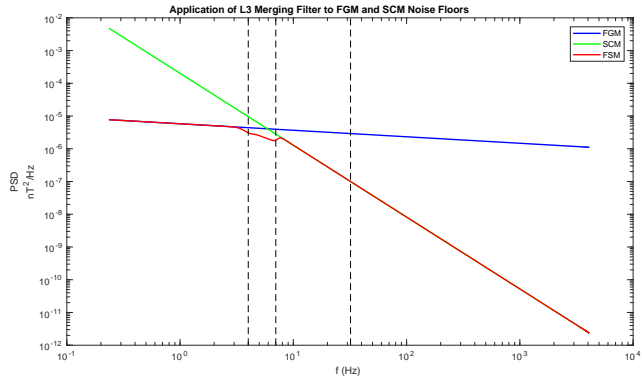}
\caption{Result of applying the merging filter to the DFG and SCM noise floors.} 
\label{fig:Filter}
\end{figure}

\section{Applications}
\label{sec:Applications}



\section{Summary}
\label{sec:Summary}

\bibliographystyle{spr-mp-nameyear-cnd}  
\bibliography{/Users/argall/Documents/library}                

%

\end{document}